\documentclass[12pt,preprint]{aastex}
\usepackage{graphicx}

\slugcomment{To be submitted to the Astronomical Journal}
\shorttitle{Close to the Dredge}
\shortauthors{Drake et al.}
\begin{document}

\title{Close to the Dredge: Precise X-ray C and N Abundances in 
$\lambda$~Andromeda and its Precocious RGB Mixing Problem}

\author{Jeremy J.~Drake\altaffilmark{1}, B. Ball\altaffilmark{1}, John J.\
  Eldridge\altaffilmark{2},  J.-U.~Ness\altaffilmark{3}, Richard J.~Stancliffe\altaffilmark{4}}

\affil{Smithsonian Astrophysical Observatory, MS-3, 60
Garden Street, Cambridge, MA 02138, USA. Email: jdrake@cfa.harvard.edu}
\affil{Institute of Astronomy, University of Cambridge, Madingley Road, Cambridge CB3 0HA, UK. Email: jje@ast.cam.ac.uk}
\affil{European Space Agency, XMM-Newton Observatory SOC, SRE-OAX, 
                 Apartado 78, 28691 Villanueva de la Ca\~nada, Madrid, Spain. Email: Jan-Uwe.Ness@sciops.esa.int}
\affil{Centre for Stellar and Planetary Astrophysics, School of Mathematical Sciences, Monash University, Melbourne, VIC 3800, Australia.  Email: richard.stancliffe@monash.edu}

\begin{abstract}

{\it Chandra} LETG+HRC-S and {\it XMM-Newton} RGS spectra of H-like C and N lines formed in the corona of
the primary star of the RS~CVn-type binary $\lambda$~And, a mildly metal-poor G8~III-IV first
ascent giant that completed dredge-up $\sim 50$~Myr ago, have
been used to make a precise measurement of its surface C/N ratio.  We
obtain the formal result [C/N]=$0.03\pm0.07$, which is typical of old
disk giants and in agreement with standard dredge-up theory for stars
$\la 1 M_\odot$.  In contrast, these stars as a group, including
$\lambda$~And, have $^{12}$C/$^{13}$C$\la 20$, which is much lower
than standard model predictions.  We show that the abundances of the old disk giants are consistent with models including thermohaline mixing that begins at the red giant branch luminosity function ``bump''.  Instead, 
$\lambda$~And indicates that the
$^{12}$C/$^{13}$C anomaly can be present immediately following dredge-up, contrary to current models of extra mixing on the red giant branch.   In the context of other recent C and N  abundance results for RS~CVn-type binaries it seems likely that the anomaly is associated with either strong magnetic activity, fast rotation, or both, rather than close binarity itself.

\end{abstract}


\section{Introduction}
\label{s:intro}

When a low-mass star ($0.8-2M_\odot$) evolves off the main sequence
and up the red giant branch (RGB), its outer convective envelope
extends inward, probing the CN-processed region of the
hydrogen-burning core and propagating the processed material up to the
stellar surface.  Standard stellar evolution models predict this
``dredge-up'' to result in a decrease of the surface $^{12}$C/$^{13}$C
and $^{12}$C/$^{14}$N ratios on the stellar surface \citep{Iben:67}.
While an extensive body of observational evidence grew 
demonstrating that such abundance changes do indeed occur, the changes
observed were often more extreme than evolutionary model
predictions \citep[see, e.g., reviews
by][]{Iben.Renzini:84,Charbonnel:05,Chaname.etal:05}.  The more
salient disagreements were for metal-poor field and globular cluster
giants, for which the $^{12}$C/$^{13}$C ratio approaches the CN-cycle
equilibrium value of $\sim 4$ \citep[e.g.\ ][]{Suntzeff.Smith:91,Shetrone.etal:93,
Shetrone:96,Carretta.etal:00,Gratton.etal:00,Keller.etal:01}.  

Since the realisation
that both $^{12}$C/$^{13}$C and $^{12}$C/$^{14}$N ratios decrease with
increasing luminosity along the RGB \citep[e.g.\
][]{Gilroy.Brown:91,Kraft:94,Charbonnel.etal:98,Gratton.etal:00,Keller.etal:01}, a variety of mechanisms to produce extra mixing between core and envelope on the RGB have been proposed to explain the  observations \citep[see, e.g.,][for a recent summary]{Angelou.etal:11}.  Of these, the thermohaline instability pointed out by 
\citet[][see also \citealt{Eggleton.etal:06} who first noted the  molecular weight inversion that causes it]{Charbonnel.Zahn:07} appears most promising.  The instability sets in 
beyond the RGB ``bump''---the point in the RGB luminosity function where the outward progress of the H-burning shell encounters the compositionally uniform layers resulting from the deepest extent of convection during the first dredge-up.  It results from the mean molecular weight of this region exceeding that of a layer just above the H-burning shell in which the mean molecular weight is reduced by the reaction $^3$He($^3$He,2p)$^4$He \citep{Abraham.Iben:70,Ulrich:71}.  


Several recent studies have discussed thermohaline mixing and its apparent success in comparisons of model predictions with observed abundances on the RGB \citep[e.g.][]{Charbonnel.Zahn:07,Recio-Blanco.De_Laverny:07, Stancliffe.etal:09,Charbonnel.Lagarde:10,Cantiello.Langer:10,Denissenkov:10,Smiljanic.etal:10,Tautvaisiene.etal:10b,Angelou.etal:11}.   Thermohaline mixing on the RGB changes both the surface $^{12}$C/$^{13}$C and $^{12}$C/$^{14}$N ratios after the end of the first dredge-up, in contrast to classical models that do not include extra mixing.  In principle, comparison of predicted and observed values of both ratios should provide a more stringent test of the theory than either ratio alone.  The $^{12}$C/$^{13}$C ratio can usually be determined from CN molecular features for RGB stars with a precision of $\sim 20$\%\ or so---useful for comparison with and constraining models.  Unfortunately, assessment of photospheric N abundances is not so precise and usually has to rely on similar CN features.  The N abundance then also depends on the derived C abundance through molecular equilibrium, and the resulting uncertainties in the C/N ratio can be quite large and rarely below 0.2~dex.  The use of field RGB stars for confronting model predictions is also hampered by their uncertain masses and evolutionary phases, especially near spectral type K0 where the evolutionary tracks of clump stars of different mass overlap with those of first ascent  stars. 

Here, we draw attention to nearby evolved active binary stars, whose masses and
evolutionary phases can generally be much better constrained than for
field stars, as offering potentially valuable laboratories for further study
of dredge-up and post dredge-up mixing.  \citet{Denissenkov.etal:06b} have also piqued interest in these stars through theoretical modelling that indicates extra mixing on the RGB might be induced by tidal spin-up.
In this context, the nearby (26~pc) old disk giant
$\lambda$~And presents an interesting case: it is a mildly metal-poor first-ascent G8~III-IV star at an evolutionary phase in which CN-cycle products should have just
appeared at its surface due to first dredge-up (\citealt{Savanov.Berdyugina:94,Donati.etal:95,Ottmann.etal:98,Tautvaisiene.etal:10}; further characteristics are listed in
Table~\ref{t:params}).  Other than its close binarity, it is similar to members of the sample of mildly metal-poor ([Fe/H]$\sim -0.5$) disk giants that \citet{Cottrell.Sneden:86} found to have  {\it unevolved}
(approximately solar) C/N ratios, but lower
$^{12}$C/$^{13}$C ratios of $\sim 10$--30 than predicted by canonical models
and reminiscent some of those seen in earlier studies of field giants toward solar metallicity
\citep{Lambert.Ries:81,Kjaergaard.etal:82}.  The unevolved C/N ratios contrasted 
with the study of slightly more massive giants in the solar metallicity cluster M67 by \citet{Brown:87}, 
which found a palpable decline of C/N on the ascent of the giant branch.  
Cottrell \& Sneden noted that the
C/N ratios in the old disk giants ``have probably remained unchanged
since their formation''.  Standard dredge-up predictions for red
giants of mass $\la 1M_\odot$ thought typical of the Cottrell \&
Sneden sample indeed indicate little change in the surface C/N ratio,
but also only mild reduction in the $^{12}$C/$^{13}$C ratio to values
$\ga 30$.  The observed low $^{12}$C/$^{13}$C ratios suggest that additional 
mixing of CN-processed material has occurred further to that of classical model
predictions, but of a nature as not to change the surface C/N ratio.
Mixing of such a characteristic was initially proposed by, e.g.,
\citet{Dearborn.Eggleton:77,Sweigart.Mengel:79,Hubbard.Dearborn:80,Lambert.Ries:81} 
in order to explain $^{13}$C rich stars such as Arcturus, which also
has [C/N]$\sim 0$.

Of the studies to date confronting thermohaline mixing model predictions with observations, little attention has been given to mildly metal-poor stars 
with masses $< 1 M_\odot$.  
In $\S$\ref{s:obs_dat} we use a precise {\em X-ray}
measurement of the surface C/N abundance of $\lambda$~And to show that it is
typical of old disk giants from the \citet{Cottrell.Sneden:86} sample.  In $\S$3, we combine this measurement with the  $^{12}$C/$^{13}$C assessments  of \citet{Tautvaisiene.etal:10} and \citet{Savanov.Berdyugina:94} and compare the results with predictions from a state-of-the-art stellar evolutionary model including thermohaline mixing.  


\section{Observations and Analysis} 
\label{s:obs_dat} 

We estimate the surface C/N abundances from emission lines of H-like ions of C and N
seen in {\it Chandra} Low Energy Transmission Grating (LETG) X-ray
spectra of $\lambda$~And.  Our analysis follows those presented
previously by \citet{Drake:03} and \citet{Drake.Sarna:03}, to which we
refer the reader for details; see also \citet{Schmitt.Ness:02}.  Our
method exploits the insensitivity of the relative intensities of the
C~VI~$\lambda 33.7$ and N~VII~$\lambda 24.7$ lines in active stars to
the coronal temperature structure: the line ratio depends essentially
only on the relative C and N abundances.

{\it Chandra} X-ray spectra of $\lambda$ And were obtained on 2002
July 22 and 23 (ObsID 2558 and 3722, respectively) in two separate
observations of 50 ks each, using the LETG and High Resolution Camera
spectroscopic (HRC-S) detector in its standard instrument
configuration.  Data were obtained from the Chandra Data
Archive\footnote{http://asc.harvard.edu/cda}, and were reduced using
the CIAO software package version 3.2.  This latter processing
included filtering of events based on observed signal pulse-heights to
reduce background.   Spectra were analysed using the PINTofALE\footnote{PINTofALE is freely
  available from http://hea-www.harvard.edu/PINTofALE/} 
IDL\footnote{Interactive
Data Language, Research Systems Inc.}  software suite
\citep{Kashyap.Drake:00}.

{\it XMM-Newton} observed $\lambda$~And on 2001 January 26 for 32~ks.  Data were processed and reduced using standard {\it XMM-Newton} Science Analysis System software, and RGS spectra were extracted with the {\sc rgsproc} task.  Spectral line fluxes were measured by fitting line profiles using the {\sc cora} program \citep{Ness.Wichmann:02}.  Measured lines intensities for both {\it Chandra} and {\it XMM-Newton} spectra are listed in Table~\ref{t:lambda_lines}.

The LETG spectrum of
$\lambda$~And for the wavelength range that includes both the N~VII
and C~VI lines is illustrated in Fig.~\ref{f:spectra}, alongside spectra of 
the single giant $\beta$ Ceti (K0 III) and the unevolved dwarf $\epsilon$
Eri (K2 V).  The latter two stars were also used by
\citet{Drake.Sarna:03} as examples of spectra of evolved and unevolved
comparison 
stars.  Simple visual inspection reveals that the $\lambda$~And
subgiant C/N line strength ratio is similar to that of $\epsilon$~Eri,
and is very unlike that of the evolved $\beta$~Ceti.
More formally, \citet{Drake:03} showed that the C/N abundance ratio by
number, $n(C)/n(N)$, is given by 
\begin{equation}
\frac{n(C)}{n(N)} = 1.85 \times \frac{(I_{C}/A_{Ceff})}{(I_{N}/A_{Neff})},
\end{equation}
where $I_{C}$ and $I_{N}$ are the number of counts in the C~VI (24.7
{\AA}) and N~VII (33.7 {\AA) lines, and $A_{Ceff}$ and $A_{Neff}$ are 
the effective area normalising factors (in cm$^{2}$) at the
appropriate wavelengths of the C and N transitions, respectively. 
Applying this formula to the
$\lambda$~And observations gives $n(C)/n(N) =3.1\pm 0.75$ and $4.3\pm
0.6$ for ObsIDs 2558 and 3722, respectively; co-adding the observations
leads to $n(C)/n(N) =3.8\pm 0.6$, or [C/N]$=-0.03\pm 0.07$ on the
\citet{Asplund.etal:09} solar abundance scale, which we adopt as our
final result.  Additional systematic
errors arising from uncertainties in instrument calibration and atomic
data are expected to be $\la 10\%$ \citep{Drake:03}.  

In light of the now well-documented chemical fractionation related to element first ionization potential (FIP) that occurs between the coronae and photospheres of stars \citep[e.g.][]{Drake:03b}, it might be questioned whether the coronal C/N ratio derived here is representative of that of the underlying photosphere.  
\citet{Drake:03} studied the coronal C/N ratio in the active binary V711~Tau and 
concluded that C and N are not fractionated to
any significant extent with respect to the photosphere.  This is 
expected based on the similar high FIPs of these elements (11.3 and 14.5 eV for C and N, respectively).  While we cannot rule out
relative fraction among C and N with absolute certainty, we therefore consider our 
derived C/N abundance ratio to be
directly applicable to the photosphere of $\lambda$~And.

Our C/N ratio is in good agreement with the earlier estimates of
[C/N]$=-0.24$ and $-0.25$ (adjusted to the \citet{Asplund.etal:09} scale) by
\citet{Savanov.Berdyugina:94} and \citet{Tautvaisiene.etal:10}, bearing in 
mind the ``0.2-0.3 dex''
uncertainty they assess for their C/N abundance ratio.

\section{Discussion} 
\label{s:discuss} 

The currently favoured solar (unevolved) $n(C)/n(N)$ abundance ratio is 3.98
(Asplund et al.\ 2009), with stated uncertainties of 0.05~dex for both
C and N.  C/N ratios for galactic field stars are known with somewhat
less precision.  Galactic disk dwarfs show [C/N]=0 down to metallicities of [Fe/H]$=-0.4$, though with some scatter attributable to uncertainties in N abundances determined from weak N~I lines \citep{Reddy.etal:03}.  \citet{Mishenina.etal:06} find no significant trend of C/N with [Fe/H] over a similar metallicity range for a sample of 177 disk giants, indicating that the post-dredge-up ratio also evolves similarly for solar metallicity and mildly metal-poor stars.

Our observed C/N ratio for
$\lambda$~And is perfectly consistent with an unevolved composition,
showing no signs of post dredge-up change.  In contrast,
\citet{Savanov.Berdyugina:94} and \citet{Tautvaisiene.etal:10} found an unambiguous signature of
dredge-up in the carbon isotope ratio for which they derived
$^{12}$C/$^{13}$C$=20\pm 5$ and $14$, respectively.  In the following we adopt the average of these values, $^{12}$C/$^{13}$C$=17\pm 5$, though our conclusions would be unchanged regardless of whether we adopted either result.  
This CN signature of $\lambda$~And is
typical of the old disk giants analysed by \cite{Cottrell.Sneden:86},
in keeping with its mild metal deficiency credential
([Fe/H]$=-0.5$ on the \citealt{Asplund.etal:09} scale;
\citealt{Savanov.Berdyugina:94,Donati.etal:95,Ottmann.etal:98,Tautvaisiene.etal:10}).  

By comparison with model evolutionary tracks, we can estimate the
evolutionary state of $\lambda$~And.  We used a version of the STARS code
\citep[][and references therein]{Eggleton:71,Pols.etal:95,Stancliffe.Eldridge:09}
with updated opacities \citep{Eldridge.Tout:04,Stancliffe.Glebbeek:08} and 
nucleosynthesis routines \citep{Stancliffe.etal:05,Stancliffe:05} to generate
evolutionary tracks from the main sequence 
for stars with metallicity $z=0.008$ and masses in the range
0.7--1.7~$M_\odot$.  Thermohaline mixing was included in these models following 
\citet{Stancliffe.etal:09}, using the prescription of \citet{Kippenhahn.etal:80}, with the diffusion coefficient multiplied by a factor of 1000, following \citet{Charbonnel.Zahn:07}.  The resulting evolutionary tracks are illustrated in
Figure~\ref{f:tracks}, together with the data point for $\lambda$~And
(see Table~\ref{t:params}).  Also shown is the locus describing the
evolutionary stage at which stars of different mass essentially reach the end of
the first dredge-up.  While dredge-up end is formally defined as the point at which the convection zone reaches its deepest penetration in mass, we defined it here as the point at which the surface $^{12}$C/$^{13}$C ratio reaches within 5\%\ of what would be the final RGB values in the absence of additional mixing processes;  after this, canonical models predict their surface abundances 
remain constant until later dredge-up phases.  We use this looser definition of the end of dredge-up to illustrate that the $^{12}$C/$^{13}$C  ratio is predicted to reach close to its final value significantly below the RGB bump.

Figure~\ref{f:tracks} shows that $\lambda$~And lies a factor of 2-3 in
luminosity below horizontal branch tracks characterising the clump, 
and so must be a first ascent giant \citep[see also the extensive discussion
  of][]{Donati.etal:95}.  It also lies just beyond the ``end'' of first dredge-up; using the STARS tracks, we estimate
that this phase was completed $\sim 50$~Myr ago, in qualitative 
agreement with the earlier assessment of 
\citet{Donati.etal:95}.  We also estimate
from these evolutionary tracks a mass for $\lambda$~And of
$1.0\pm0.2M_\odot$, which compares favourably with the spectroscopic
estimate based on $\log g$ and the radius (Table~\ref{t:params}) of
$1.3^{+1.0}_{-0.6}M_\odot$. 
This is consistent, within experimental error, with the estimate of
\citet{Donati.etal:95}; our slightly larger value is mostly due to our
adopted surface gravity being higher \citep[see analyses
by][]{Savanov.Berdyugina:94,Ottmann.etal:98,Tautvaisiene.etal:10}.

The value of the precise X-ray measurement of C/N is apparent in
Figure~\ref{f:c12c13cn}, which compares the observed C isotope and C/N
ratios for $\lambda$~And with predictions from canonical evolution models past the first dredge-up but with no additional post-dredge-up mixing 
\citep{Schaller.etal:92,VandenBerg:92,Girardi.etal:00}.   
In accordance with well-established dredge-up 
results, models predict a reduction in 
$^{12}$C/$^{13}$C with increasing mass, with only very mild reductions
for the lowest mass stars.  
Also shown are the observed ratios for the old disk giants from
\citet{Cottrell.Sneden:86}.  This kinematically-selected sample comprises low mass stars with $M \la 1M_\odot$,  very much like $\lambda$~And.  
$\lambda$~And and the old disk giants have
C/N ratios more or less in agreement with the models, but have much lower
$^{12}$C/$^{13}$C ratios.  There is some selection effect in the
latter group because stars with higher $^{12}$C/$^{13}$C ratios tend
to be represented only by lower limits, though this applies only to 4
out of the 34 stars in the \citet{Cottrell.Sneden:86} sample; no
$^{12}$C/$^{13}$C data are available for a further 8 stars.
Statistically, the majority of the \citet{Cottrell.Sneden:86} sample
are expected to be core He-burning stars, since this evolutionary
phase lasts several times longer than the first RGB ascent.  
The problem with the abundance ratios for these old disk giants is that they exhibit much more extreme processing of $^{12}$C to $^{13}$C than the models, but essentially no processing of $^{13}$C to $^{14}$N.  

The STARS models with thermohaline mixing predict changes in the surface  $^{12}$C/$^{13}$C and C/N ratios above the RGB `bump'.  The $^{12}$C/$^{13}$C vs.\ C/N loci from the main-sequence to the RGB tip for four models corresponding to different stellar masses are illustrated in Figure~\ref{f:thermohaline}.   The most relevant of these for the sample under consideration are the 0.7 and $1.0 M_\odot$ ones.  While the Cottrell \& Sneden sample has systematically higher C/N ratios on average than the models by about 50\%,  a systematic error of this magnitude in the observed values probably cannot be ruled out given the 0.2~dex uncertainty on each.  The range of observed $^{12}$C/$^{13}$C ratios is instead in very good agreement with the thermohaline mixing predictions.  

At face value, the generally reasonable agreement between observed and model abundances for the metal-poor old disk star sample is encouraging.  However,  $\lambda$~And does present a problem.  Its C/N ratio is significantly higher than model predictions, though for a single object such a discrepancy might be  attributed to cosmic variance.  More important is the evolutionary phase at which the thermohaline mixing begins to affect the surface $^{12}$C/$^{13}$C ratio.  This is above the RGB bump.   \citet{Tautvaisiene.etal:10} pointed out that $\lambda$~And instead lies well below the bump, as demonstrated in Figure~\ref{f:tracks}, and indicates that $^{12}$C/$^{13}$C is anomalous earlier in its evolution, and perhaps immediately following dredge-up.  

One possible source of the anomaly is contamination from its companion.  There is no direct observational information to assess with certainty the nature of the companion, though \citet{Donati.etal:95} presented cogent arguments based on the
asynchronicity of the rotation and orbital periods of $\lambda$~And to
rule out a white dwarf nature for the secondary.  Such a star might
otherwise have contaminated the current primary with carbon-rich
material during its asymptotic giant branch (AGB) phase.    
However, finding a companion that is able to provide material of a suitable composition is difficult.  Low-mass AGB stars that undergo the third dredge-up become rich in carbon-12, yet do not produce much in the way of carbon-13 and nitrogen-14. For example, \citet{Karakas.Lattanzio:07} find that a $1.5 M_\odot$ star of $Z=0.004$ produces ejecta with a $^{12}$C/$^{13}$C ratio of over 100 and a C/N ratio of about 10.  \citet{Stancliffe.Jeffery:07} find similar values for the ejecta of a $1.5M_\odot$ star of $Z=0.008$. These values are too high to match those observed in $\lambda$~And, even if one allows for the dilution of any accreted material via mixing with the pristine material of the receiving star.  One may wish to invoke some sort of extra mixing processes at work during the AGB in order to lower these values.  The physical nature of this extra mixing process is unknown.  It is unlikely to be thermohaline mixing, as this has been shown unable to produce $^{12}$C/$^{13}$C ratios as low as 10 at lower metallicity than that of $\lambda$~And even though it is more effective at low metallicity \citep{Stancliffe:10}. There is also debate about whether this is actually required in low-mass AGB stars (see \citealt{Busso.etal:10} and \citealt{Karakas.etal:10} for opposite sides of this debate). Appealing to a higher mass companion, one that underwent hot bottom burning on the AGB, is also unlikely to help: the models of \citet{Karakas.Lattanzio:07} suggest that these would produce $^{12}$C/$^{14}$N less than 1---too low to match the observations.

We also note here a fundamental difference between $\lambda$~And and the subgiants in the globular clusters NGC~6752 and 47~Tuc that were found to have $^{12}$C/$^{13}$C$\la 10$ by  \citet{Carretta.etal:05}: unlike $\lambda$~And these stars also have depleted C and enhanced N abundances.  \citet{Carretta.etal:05} interpreted these abundances as the signature of contamination by mass loss from intermediate-mass AGB stars.  

Recently, the C isotope and C/N ratios have been estimated for two additional RS~CVn-type binaries, 29~Dra and 33~Psc.  \citet{Barisevicius.etal:10,Barisevicius.etal:11} find similar C/N ratios ([C/N]$\sim-0.25$) to the \citet{Tautvaisiene.etal:10} value for $\lambda$~And, but find $^{12}$C/$^{13}$C$=16$ for 29~Dra and $30$ for 33~Psc.  The latter is ``normal'' for a post-dredge-up star, and indeed 33~Psc, like $\lambda$~And, lies below the bump luminosity \citep{Barisevicius.etal:11} and is not expected to have experienced additional post-dredge-up mixing episodes.  The isotope ratio for 29~Dra is instead similar to that of $\lambda$~And.  Based on the atmospheric parameters of  \citet{Barisevicius.etal:10}, 29~Dra is probably cooler by $\sim 100$~K and more luminous by 0.05~dex than $\lambda$~And, and therefore lies slightly above and to the right of $\lambda$~And in Figure~\ref{f:tracks}.  Nevertheless, it is likely still located just below the  luminosity bump and would therefore appear anomalous in terms of the $^{12}$C/$^{13}$C ratio.

$\lambda$~And, and likely 29~Dra, then seem to present a challenge to the current description of  thermohaline mixing as the controlling factor in post-dredge-up  surface abundance changes.  While this mechanism appears generally successful in explaining post-dredge-up CNO and light element surface abundances  ($\S$\ref{s:intro}), the diffusion coefficient dictating the rate of mixing is not well-constrained \citep{Charbonnel.Zahn:07,Denissenkov:10}.  Three-dimensional numerical simulations of thermohaline convection by \citet{Denissenkov.Merryfield:11} and \citet{Traxler.etal:11} find a mixing rate that is at least a factor of 50 lower than that required by models to match observed RGB abundance changes. 
In essence, the penetrating structures of one fluid into the other that effect the mixing, commonly referred to as ``salt fingers'', have a length-to-width aspect ratio in the simulations that is too small.  
Both studies concluded that thermohaline convection driven by $^3$He burning is unlikely to be the sole mechanism of extra mixing on the RGB and must be enhanced or augmented by other processes.  Interestingly for the magnetically-active $\lambda$~And, \citet{Denissenkov.Merryfield:11} suggested strong toroidal magnetic fields arising from differential rotation in the radiative zone might enable more growth of thicker fingers.  Alternatively,  \citet{Charbonnel.Zahn:07b} note that strong magnetic fields can act to damp the thermohaline instability and suggest this as a means of suppressing extra mixing in the RGB descendents of Ap stars.

Even if the magnetic activity of $\lambda$~And might provide some assistance to the process, it does not solve the problem that its surface abundances are changed before the predicted on-set of the thermohaline instability.  Rotation-driven mixing, once thought the main abundance-modifying mechanism on the RGB,  also would not help for similar reasons, although \citet{Palacios.etal:06} have essentially ruled the process out for producing significant surface abundance modifications.  One possible alternative is the magnetic mixing process proposed by \citet{Busso.etal:07}, in which buoyant flux tubes generated in a stellar dynamo operating near the hydrogen-burning shell transport processed matter upward into the convective envelope. Again, the challenge for such a mechanism to work for $\lambda$~And would be to bring sufficient $^{13}$C-rich material to the surface so soon after dredge-up.

We are otherwise drawn back to the discussion of the
$^{13}$C-rich giants by \citet{Lambert.Ries:81}, who reasoned that
some $^{13}$C must be removed from the CN-processing zone prior to
conversion through proton capture to $^{14}$N on the main sequence
(see also, e.g.\ \citealt{Dearborn.Eggleton:77}, \citealt{Sweigart.Mengel:79}, and the 
``magnetic mixing'' of \citealt{Hubbard.Dearborn:80}).  Thermohaline mixing seemed to solve much that was raised in that discussion, though in the light of the conclusions of \citet{Denissenkov.Merryfield:11}, the odd exception such as $\lambda$~And and 29~Dra also seems to point to other effects.   Existing studies of rotationally-driven mixing and other mechanisms do not yet generally include aspects such as close binarity and strong magnetic fields that distinguish $\lambda$~And from other stars in RGB abundance surveys.   Theoretical models including tidal interaction and spin-up of giants in RS~CVn-type binaries developed by \citet{Denissenkov.etal:06b} present an interesting exception.  

\citet{Denissenkov.etal:06b} found evolutionary models with tidal spin-up caused by a close companion to exhibit large excursions of $M_V \sim 0.7$--0.8 at the luminosity bump, and evolution through this phase consequently took 50\%\ longer than for single stars.  They cited this as a possible explanation for the propensity of known tidally-interacting binaries to comprise one component on the lower part of the RGB.  The models assumed synchronization of orbital and rotation periods had been established. $\lambda$~And is notoriously unsynchronized, but it lies only just below the bump luminosity.  Were such large bump excursions to occur, they would allow $\lambda$~And and 29~Dra to be``post-bump''  stars, lying on the downward lower luminosity excursion.   The extended stay at this point in its evolution would also provide greater opportunity for mixing of $^{13}$C-rich material into its convection zone by rotation, thermohaline or other mixing process.

The question then arises why $\lambda$~And and 29~Dra exhibit C isotope ratio anomalies but 33~Psc does not?  \citet{Barisevicius.etal:11} pointed to the low activity of 33~Psc compared to that of $\lambda$~And and 29~Dra.  The orbital period of 33~Psc is 79.4 days \citep{Harper:26} though its rotation velocity is not constrained.  The orbital period of 29~Dra is much longer, 904 days, but it has a rotation period of about 30 days \citep{Hall.etal:82,Zboril.Messina:09}, which is shorter than that of $\lambda$~And (54 days), and based on activity indices is likely a faster rotator than 33~Psc.  Since 29~Dra is a much wider binary than either $\lambda$~And or 33~Psc, tidal interactions would appear not to be the primary factor governing the isotope anomaly, and instead rotation and activity appear to be the culprits. 



\section{Conclusions}

A precise X-ray measurement of H-like C and N formed in the corona of
the G8~III-IV primary of the RS~CVn-type binary $\lambda$~And has revealed a solar C/N
abundance ratio that is essentially unchanged since the star was
formed.  This is in qualitative agreement with evolutionary
calculations for such low-mass stars ($\la 0.9M_\odot$), but is quite
inconsistent with its photospheric ratio $^{12}$C/$^{13}$C$=14$
\citep{Savanov.Berdyugina:94,Tautvaisiene.etal:10} that is much lower than predicted
by theory.  This abundance pattern is typical of the old disk giants
first brought to prominence by \citet{Cottrell.Sneden:86}.
$\lambda$~And is a first ascent red giant that underwent dredge-up
only $\sim 50$~Myr ago; its anomalously low
$^{12}$C/$^{13}$C has appeared immediately after dredge-up.
We echo the conclusions of \citet{Tautvaisiene.etal:10} that extra mixing on the RGB in magnetically-active low-mass stars like $\lambda$~And and 29~Dra appears to act below the luminosity function bump, in contradiction with current thermohaline mixing models.  Magnetic flux tube buoyant mixing \citep{Busso.etal:07} would appear to warrant more detailed investigation.  Evolutionary models of tidally-interacting binaries by \citet{Denissenkov.etal:06b} that predict large luminosity excursions at the bump might also allow $\lambda$~And to be a ``post-bump'' giant and alleviate its otherwise precociously diminished C isotope ratio.  Such a mechanism would appear to be less promising for wider binaries such as 29~Dra.

\acknowledgments

We thank the
NASA AISRP for providing financial assistance for the development of
the PINTofALE package.  JJD was supported by NASA contract NAS8-39073
to the {\em Chandra X-ray Center} during the course of this research.
WB was supported by {\it Chandra} award number AR4-5002X issued by the
{\it Chandra} X-ray Center.  JJD thanks H.~Tananbaum and the CXC science team for advice and support.


\newpage


\begin{deluxetable}{lccccccccc}  
\tabletypesize{\scriptsize}
\tablecaption{Summary of parameters of the $\lambda$ And primary}
\tablewidth{0pt}
\tablehead{
\colhead{Spec.\ Type}
&\colhead{Dist (pc)\tablenotemark{a}}
&\colhead{P$_{rot}$ (d)\tablenotemark{b}}
&\colhead{$T_{eff}$ (K)\tablenotemark{c}}
&\colhead{$\log g$\tablenotemark{c}}
&\colhead{[Fe/H]\tablenotemark{c}}
&\colhead{$M\, (M_\odot)$\tablenotemark{d}}
&\colhead{$R\, (R_\odot)$\tablenotemark{e}}
&\colhead{$\log L/L_\odot$\tablenotemark{f}}
&\colhead{$L_{X}$ (erg s$^{-1}$)\tablenotemark{g}}
}
\startdata
  G8III-IV+ & 25.8 & 54 & $4800\pm 100$ & $2.75\pm 0.25$ & $-0.5$ & 
  $1.3^{+1.0}_{-0.6}$ & $7.0\pm 0.7$ &  
  $1.37\pm 0.04$ & $2.95\times 10^{30}$
\enddata
\tablenotetext{a}{{\it Hipparcos} distance from \citet{van_Leeuwen:07} .}
\tablenotetext{b}{\citet{Landis.etal:78}}.
\tablenotetext{c}{Adopted here based on estimates from
  \citet{Savanov.Berdyugina:94,Donati.etal:95,Ottmann.etal:98,Soubiran.etal:08,Tautvaisiene.etal:10}; [Fe/H]
is expressed relative to the solar composition $Fe/H=7.50$ \citep{Grevesse.Sauval:98,Asplund.etal:09}.} 
\tablenotetext{d}{Spectroscopic mass from $L_{bol}$, $T_{eff}$ and
  $\log g$.}
\tablenotetext{e}{Spectroscopic radius based on $L_{bol}$ and 
  $T_{eff}$.}
\tablenotetext{f}{\citet{Tautvaisiene.etal:10}; uncertainty estimated here based on the \citet{Alonso.etal:99} bolometric correction change for a 100~K temperature uncertainty at our adopted $T_{eff}$.}
\tablenotetext{g}{\citet{Dempsey.etal:93}, based on {\it ROSAT}
All-Sky Survey observations, updated to the {\it Hipparcos} distance of \citet{van_Leeuwen:07}.}
\label{t:params}
\end{deluxetable}

%

\begin{deluxetable}{lcccccccc}
\tabletypesize{\footnotesize}
\tablecaption{Measured line intensities
(in counts) for $\lambda$~And}
\tablehead{\colhead{} & \colhead{}  & \multicolumn{3}{c}{Chandra} & \colhead{} & \multicolumn{2}{c}{XMM-Newton}  
& \colhead{} \\ 
\cline{3-5} \cline{7-8} \\
\colhead{$\lambda$ (\AA)} &
\colhead{Ion} & 
\colhead{ObsID 2558}  & 
\colhead{ObsID 3722}  & 
\colhead{$A_{eff}$} &
\colhead{} &
\colhead{RGS2} & 
\colhead{$A_{eff}$} &
\colhead{Transition}
}
\startdata
24.779 & N  VII  &163.1$\pm{34}$ & 211.7$\pm{36}$ & 15.2 & & $136.1\pm 19$ & 46.7 &
{(2p) 2P$_{3/2} \rightarrow$ (1s) 2S$_{1/2}$}\\
33.734 & C  VI   &205.5$\pm{27}$ & 376.1$\pm{35}$ & 11.6 & & $138.7\pm 18$ & 22.6 &
{(2p) 2P$_{3/2} \rightarrow$ (1s) 2S$_{1/2}$}\\
\enddata
\label{t:lambda_lines}
\end{deluxetable}


\begin{figure}
\plotone{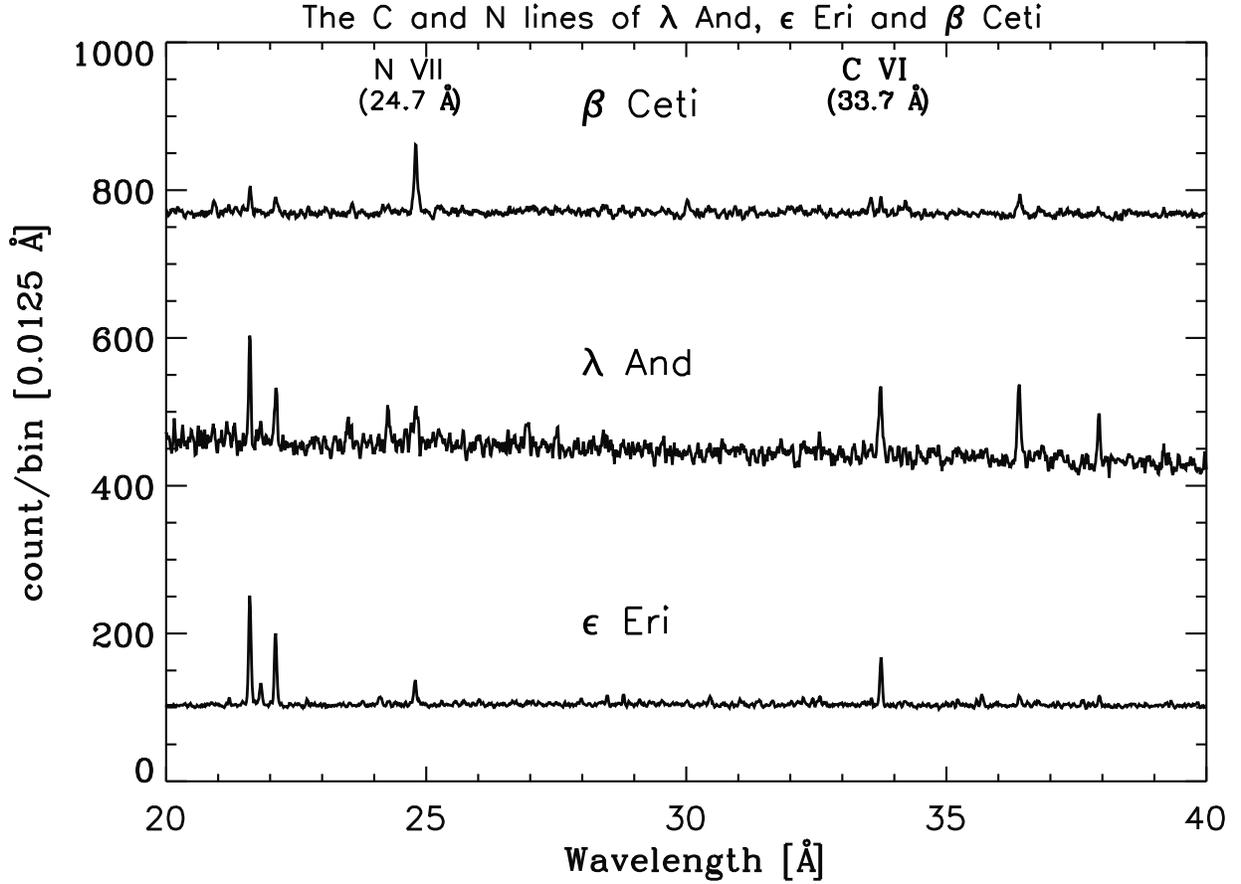}
\caption{The $\lambda$~And LETGS spectrum in the 20-40~\AA\ region,
  showing C and N lines, compared with
$\beta$~Ceti (evolved) and $\epsilon$ Eri (unevolved). The
$\lambda$ And spectrum has been multiplied by 3 for clarity.  The
  $\lambda$~And C and N line strengths are in a similar ratio to that
  of the unevolved dwarf $\epsilon$ Eri and show no significant signs
  of the C depletion evident in the spectrum of $\beta$~Ceti.
}
\label{f:spectra}
\end{figure}

\begin{figure}
\plotone{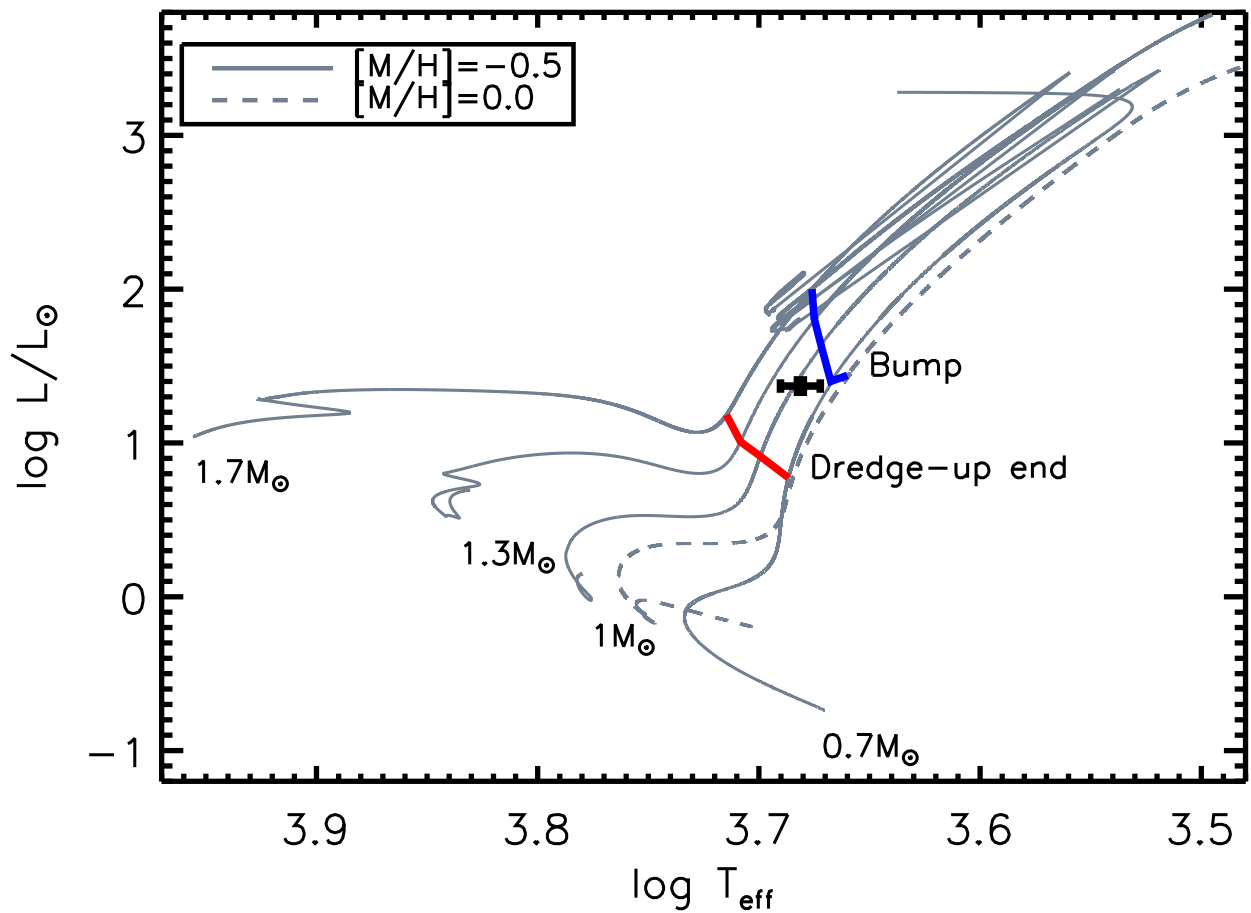}
\caption{Comparison of the effective temperature and luminosity of
  $\lambda$~And with evolutionary tracks computed using the STARS
  program for metallicity $Z=0.008$ (see text). ``Dredge-up end''
  indicates the point at which the surface abundances are within 5\%\
  of their final post-dredge-up values.  $\lambda$~And finished
  dredge-up about 50~Myr after this point.
}
\label{f:tracks}
\end{figure}

\begin{figure}
\plotone{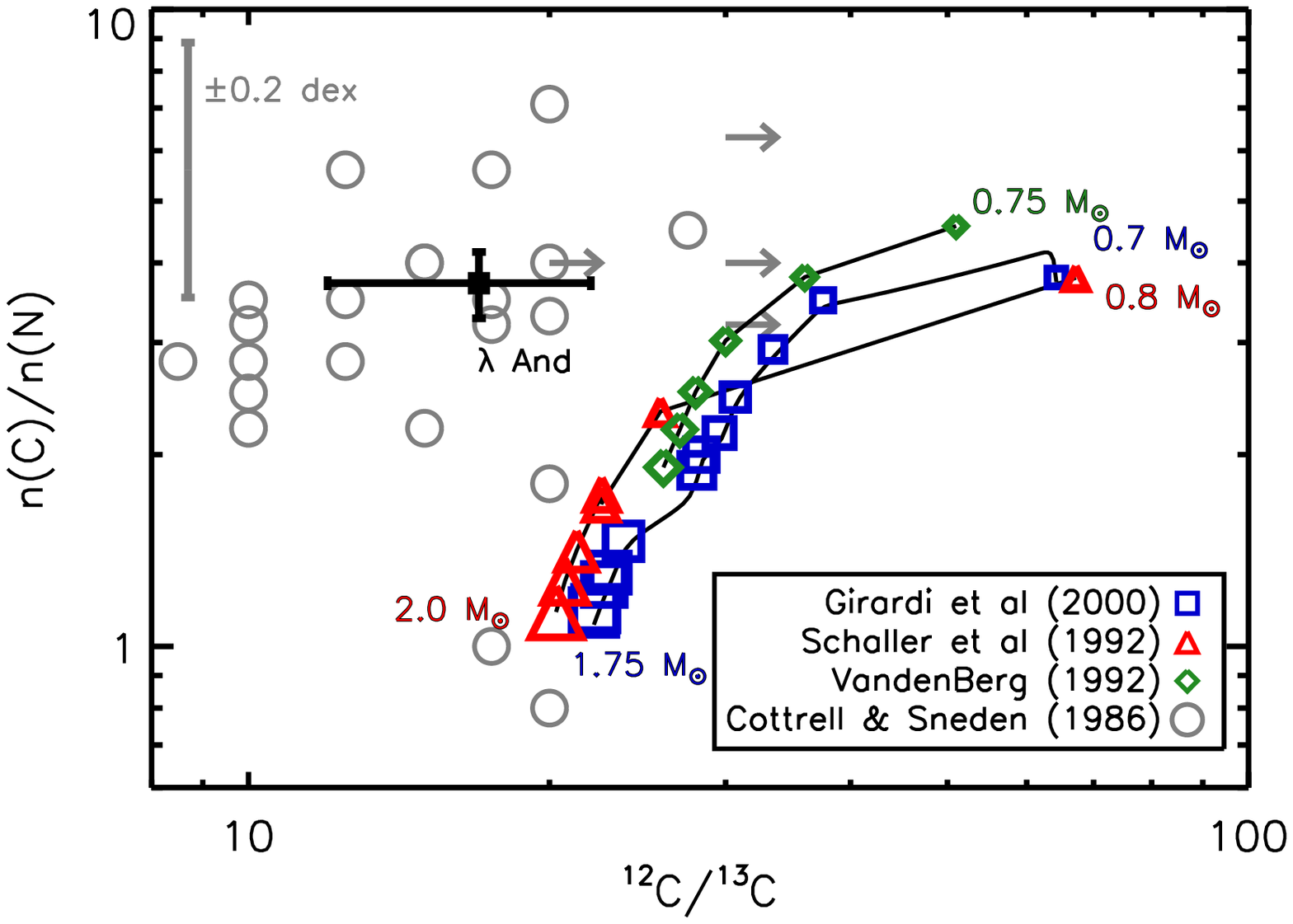}
\caption{Model predictions of the post-dredge-up surface C/N abundance
  ratio (by number) as a function of the $^{12}$C/$^{13}$C ratio for
  stars of different mass (denoted by symbol size), from
  \citet{Schaller.etal:92,VandenBerg:92} and
  \citet{Girardi.etal:00}.  Also shown are the 
  $^{12}$C/$^{13}$C and C/N ratios estimated for old disk giants by
  \citet{Cottrell.Sneden:86}, together with a representative 0.2~dex
  error bar.  The X-ray $\lambda$~And C/N ratio is illustrated
  together with the $^{12}$C/$^{13}$C ratio from \citet{Tautvaisiene.etal:10}.
}
\label{f:c12c13cn}
\end{figure}

\begin{figure}
\plotone{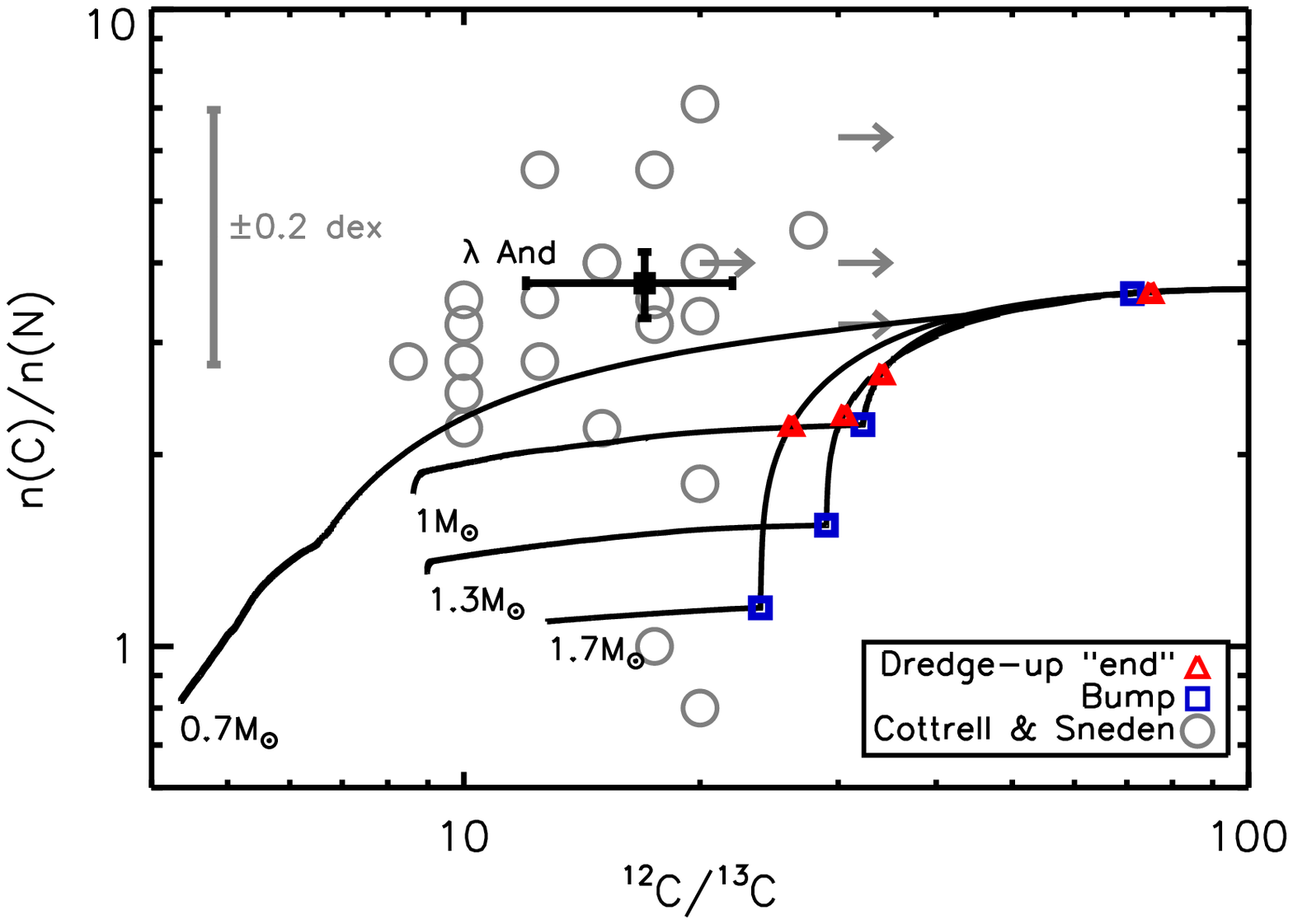}
\caption{Similar to Figure~\ref{f:c12c13cn} but illustrating STARS predictions of the post-dredge-up surface 
$^{12}$C/$^{13}$C vs C/N trajectories of models including thermohaline mixing
on the RGB compared with the ratios estimated for old disk giants by
  \citet{Cottrell.Sneden:86}.  The X-ray $\lambda$~And C/N ratio is illustrated
  together with the $^{12}$C/$^{13}$C ratio from \citet{Tautvaisiene.etal:10}.   Evolution from the main-sequence toward the tip of the RGB corresponds to progress along the loci from right to left.  
}
\label{f:thermohaline}
\end{figure}

\end{document}